\documentclass[a4paper]{jpconf}
\usepackage{graphicx}
\begin{document}
\title{When the clock strikes: Modeling the relation between
circadian rhythms and cardiac arrhythmias}

\author{Pavithraa Seenivasan$^{1,2}$, Shakti N. Menon$^1$, S.
Sridhar$^{1,3}$ and Sitabhra Sinha$^1$}

\address{$^1$ The Institute of Mathematical Sciences, CIT Campus, 
Taramani, Chennai 600 113, India}
\address{$^2$ Molecular Biophysics Unit, Indian Institute of Science,
Bangalore 560 012, India}
\address{$^3$ Department of Physics and Astronomy, Ghent University, St. Pietersnieuwstraat 33, 9000 Ghent, Belgium}

\ead{sitabhra@imsc.res.in}

\begin{abstract}
It has recently been observed that the occurrence of sudden cardiac death
has a close statistical relationship with the time of day, viz., 
ventricular fibrillation is most likely to occur between
$12$am$-$$6$am, with $6$pm$-$$12$am being the next most likely period. 
Consequently there has been significant interest in understanding how
cardiac activity is influenced by the circadian
clock, i.e., temporal oscillations in physiological activity with a
period close to $24$ hours and synchronized with the day-night cycle.
Although studies have identified the genetic basis of circadian
rhythms
at the intracellular level, the mechanisms by which they influence
cardiac pathologies are not yet fully understood.
Evidence has suggested that diurnal variations in the conductance
properties of ion channel proteins that govern the excitation dynamics of 
cardiac cells may provide the crucial link. 
In this paper, we investigate the relationship between the circadian
rhythm as manifested in modulations of ion channel properties and the
susceptibility to cardiac arrhythmias by
using a mathematical model that describes electrical activity in
ventricular tissue. We show that changes in the channel conductance
that lead to extreme values for the duration of action potentials in cardiac
cells can result either in abnormally high-frequency reentrant
activity or spontaneous conduction block of excitation waves.
Both phenomena increase the likelihood of wavebreaks that are known to
initiate potentially life-threatening arrhythmias. Thus, disruptive
cardiac excitation dynamics are most likely to occur in time-intervals
of the day-night cycle during which the channel properties are
closest to these extreme values, providing
an intriguing relation between circadian rhythms and cardiac
pathologies.
\end{abstract}

\begin{quote}
``And now it is almost midnight, the moment when the page of the night
turns over into day. Almost midnight, the hour when the figure of
Death strikes the golden bell of the clock. And what will happen when
the clock strikes?''
\end{quote}
\begin{flushright}
{\em When the Clock Strikes} by Tanith Lee (1981)
\end{flushright}

%=====================================================================================================================

%=====================================================================================================================%
\section{Introduction}
Heart disease constitutes a significant public health burden worldwide, being the
leading cause of death in most developed countries - for instance, being responsible
for about $25\%$ of all deaths in the United States~\cite{Hennekens1998,Heron2013}.
A large fraction of these can be classified as {\em sudden cardiac
death}, occurring as a result of
certain types of {\em cardiac arrhythmia}, i.e., disturbances in the normal rhythmic activity of the heart that severely
impair the normal functioning of the organ. 
There are many factors that determine the occurrence of cardiac
pathologies. While ageing results in a progressive decline in the
functioning of cardiovascular systems in general~\cite{Hatch2011},
genetics~\cite{LaraPezzi2015} and lifestyle
choices~\cite{Mozaffarin2008} can result in a higher predisposition to
arrhythmias. Apart from these, the role of circadian rhythms - which
refer to temporal variations in physiological processes having a period
of approximately 24 hours - in
initiating cardiac dysfunction has become the focus of recent clinical
interest.
Circadian rhythms help in coordinating the
behavior of organisms with the terrestrial day-night
cycle. Such biological clocks in the body are realized by a set of proteins
that enable the generation of rhythmic activation through
self-sustained, transcriptional positive and negative feedback
loops~\cite{Dunlap1999,Edery2000}. Experimentally, it has been
observed
that the circadian rhythm is not driven by the environment but rather
is an intrinsic property of the organism~\cite{Edery2000}. This is
supported by the observation that these rhythms persist even in systems
maintained in vitro~\cite{Balsalobre1998,Durgan2005}. It has been postulated
that an internal clock mechanism enables the organisms to anticipate
temporal changes in their environment, thereby allowing biological
processes to occur at a favorable time in the
day~\cite{Pittendrigh1993,Ouyang1998,Paranjpe2005}. Circadian clocks
are entrained to the external day-night cycle by environmental factors
({\it zeitgebers}) that influence the timing of the molecular
mechanism~\cite{Sharma2005}.

In this paper, we have used computational modeling to investigate the relation between
onset of arrhythmias in the heart and the circadian rhythm that have been revealed by
recent experimental studies.
Over the last few decades, sophisticated models of electrical activity in cardiac cells
(that lead to mechanical contraction and hence the pumping action of the heart)
have been developed with the aim of reproducing experimentally
observed phenomenon. The focus of such modelling efforts has been to
capture relevant details from the level of the cell up to that of the
whole heart. Increased availability of experimental data and vastly
improved computational power has enabled the development of these
detailed models of cardiac activity for studying the normal and
pathological functioning of the heart. We use such a biologically realistic model
to simulate the dynamics of cardiac tissue in order to understand the
higher propensity for the onset of arrhythmias at certain times of the day-night cycle.
In particular we show that the diurnal variations that cause ion channel conductances to take extreme values can
result in significant changes in the repolarization properties of the excitation waves. This can lead to either
abnormally rapid high-frequency activity or spontaneous conduction block of propagating excitation - both
of which increase the likelihood of wavebreaks that can initiate potentially life-threatening arrhythmia.
In the next section we describe the biological phenomena underlying circadian rhythm and its 
possible relation to onset of arrhythmia. Following this we describe
the mathematical model that we have used 
for our computational investigation of excitation dynamics in simulated cardiac tissue. In the subsequent section
we describe the results of our investigation into the link between
circadian rhythms and the genesis of cardiac arrhythmias. We 
conclude with a brief summary of our results. 

\section{Circadian Rhythms}
The mammalian circadian clock involves an interaction between three
negative loops and one positive loop, executing a series of
transcriptional, translational and post-translational
events~\cite{Gachon2004}. The \textit{clock} and \textit{bmal1} genes,
which
form the preliminary units of the clock system, encode the proteins
CLOCK and BMAL1. These proteins, on hetero-dimerization, recognize the
promoter regions of the genes, \textit{per} and \textit{cry}, and
transcribe them in the nucleus \cite{Gekakis1998, Hogenesch1998}. The
\textit{per} and \textit{cry} messenger RNAs get translated in the
cytosol to form proteins, PER and CRY. These proteins, on
hetero-dimerization \cite{Kloss1998}, translocate back into the
nucleus and inhibit the activation of CLOCK/BMAL1. This, in turn,
inhibits CLOCK/BMAL1-mediated transcription, thereby decreasing the
expression of \textit{per} and \textit{cry} genes, ultimately
relieving the inhibition on CLOCK/BMAL1 \cite{Curtis2004,
Etchegaray2003, Gachon2004}. This mechanism constitutes the dominant
negative transcriptional loop of the mammalian circadian clock at the
molecular level. On the other hand, the positive transcriptional loop
of the circadian clock results from CLOCK/BMAL1 and/or PER2-mediated
induction of \textit{bmal1} expression \cite{Shearman2000}.

Depending on the cell type, mammalian circadian clocks are divided
into two major classes: the central clock located within the
suprachiasmatic nucleus (SCN) of the brain, and peripheral clocks
located in all non-SCN cells of the organism \cite{Cermakian2000,
Hirota2004}. \textit{Zeitgebers}, the factors that reset or entrain circadian
clocks, are different for the two classes of clocks. The SCN is reset
by light (through electrical signals transmitted along the
retino-hypothalamic tract) from the environment, whereas peripheral
clocks are reset by various neuro-humoral factors, specific to it
\cite{Berson2003, Brown2002}. For example, Norepinephrine
\cite{Durgan2005} and Vasoactive intestinal peptide (VIP)
\cite{Schroeder2011} have been identified to be the neuro-humoral
factors that entrain the cardiomyocyte peripheral clock to the SCN.

%--------------------------------------------------------------------------------------------------------------------
\subsection{Sudden cardiac death and the circadian clock}

The incidence of sudden cardiac death has been observed to exhibit a
diurnal variation with peaks occurring in the early morning and late
evening hours \cite{Muller1987}. The circadian expression of clock
genes and clock-controlled genes (genes that are transcribed by the
clock genes) in the heart have been recently studied with the aim of
understanding this variation \cite{Jeyaraj2012, Schroder2013}. One
such clock controlled gene observed to exhibit endogenous circadian
rhythmicity in the heart is \textit{klf15}~\cite{Jeyaraj2012}. KChIP2,
the regulatory $\beta$~subunit for the repolarizing transient outward
K$^{+}$ current was identified to be the transcriptional target of
\textit{klf15}. That is, \textit{klf15} transcribes the gene
\textit{kcnip2} that encodes the potassium ion channel protein,
KChIP2. Experimentally, it has been observed that myocardial
repolarization and the expression of ion channels exhibit endogenous
circadian rhythmicity. In \textit{klf15}-null mice, the QTc interval
(an index of myocardial repolarization) is prolonged and the $I_{\rm
to}$ (outward transient K$^{+}$ current density) reduced. In contrast,
the QTc interval shortens and $I_{\rm to}$ increases in the case of
transgenic mice with over-expressed \textit{klf15}.
This implies that \textit{klf15}-dependent transcriptional regulation
of rhythmic KChIP2 expression plays a key role in the rhythmic
variation of myocardial repolarization. A similar observation has been
made with respect to the sodium ion channel proteins
\cite{Schroder2013}. The clock gene \textit{bmal1} has been found to
directly regulate the circadian oscillation of \textit{scn5$\alpha$}
gene that encodes Nav1.5, the principal voltage-gated Na$^+$ ion
channel. Ventricular myocytes isolated from \textit{bmal1}-deleted
mutants have less Na$^+$ current compared to the ones isolated from
\textit{bmal1}-expressed mice. This suggests that the expression of
\textit{scn5$\alpha$} is a significant factor that influences the
susceptibility to cardiac arrhythmias.
To summarize, the susceptibility to cardiac arrhythmias is increased
when the temporal variation (circadian variation) in myocardial
repolarization and ion channel expression is
impaired. This is reflected in the observation of lengthened/shortened
QT interval downstream of a disruption to the circadian genes.

%=====================================================================================================================%
\section{The computational model}
Electrical activity in cardiac muscle is typically described
mathematically by a generic class of models known as excitable systems. 
An excitable medium is characterized by a stable resting state, a
metastable excited state, and a threshold which needs to be exceeded
for the system to be excited. A supra-threshold stimulus gives rise to
an excitation which is followed by a period of slow recovery (referred
to as the refractory period) during which the system cannot be re-excited.
This refractory property results in the annihilation of waves when
they collide with each other. This implies that a wavefront, on
encountering a region that has not yet completely recovered, develops
wave breaks that can give rise to spatial patterns such as spiral
waves of excitation~\cite{Jalife1999}.

In this paper, we have used a modified version of the Luo-Rudy I (LRI) model 
describing the electrical activity of cardiac
myocytes in the guinea pig ventricle~\cite{Luo1991}. The model is based on the Hodgkin-Huxley
formalism \cite{Hodgkin1952} developed for describing the action
potential of a squid giant axon. The generic form of such models are
described by a partial differential equation for the transmembrane
potential $V$:
\[\frac{\partial V}{\partial t} + \frac{I_{\rm ion}}{C} = D\, \nabla^2
V, \]
where $C$ is the membrane capacitance, $D$ is the diffusion constant
and $I_{\rm ion}$ is the total ionic current density. In the Luo-Rudy
I model, the total ionic current density is the sum of six components:
\[
I_{\rm ion} = I_{\rm Na} + I_{\rm K} + I_{\rm K1} + I_{\rm Kp} + I_{\rm Ca} + I_{\rm b}\,,
\]
where $I_{\rm Na} = G_{\rm Na}\,m^{3}\,h\,j\,(V - 54.4)$ is the fast
inward Na$^{+}$ current, $I_{\rm K} = G_{\rm K}\,x\,x1\,(V + 77.62)$,
$I_{\rm K1} = G_{\rm K1}\,K1_{\infty}\,(V + 87.95)$, and
$I_{\rm Kp} = 0.0183\,K_{\rm p}\,(V + 87.95)$ are respectively the
time-varying, the time-invariant and the plateau K$^{+}$ currents,
$I_{\rm Ca}$ = $G_{\rm Ca}\,d\,f\,(V - E_{\rm
Ca})$ is the slow inward Ca$^{2+}$ current where $E_{\rm Ca} = 7.7 -
13.0287$ ln([Ca$^{2+}$]$_{i}$) is the reversal potential dependent on
the intracellular ion concentration [Ca$^{2+}$], and $I_{\rm b} = 0.03921\,(V +
59.87)$ is the background current (leakage current). The currents are
determined by ion channel gating variables $m$, $h$, $d$, $f$ and $x$,
whose time evolution is governed by ordinary differential equations of
the form:
\[\frac{\partial \varepsilon}{\partial t} = \frac{\varepsilon_{\infty}
-\varepsilon}{\tau_{\varepsilon}}, \]
where $\varepsilon_{\infty} = \frac{\alpha\, \varepsilon}{\alpha\,
\varepsilon + \beta\, \varepsilon}$ is the steady state of
$\varepsilon$, $\tau_{\varepsilon} = \frac{1}{\alpha\, \varepsilon
+ \beta\, \varepsilon}$ is the corresponding time constant and
$\alpha\, \varepsilon$ and $\beta\, \varepsilon$ are voltage dependent
rate constants, all obtained by fitting experimental data. The parameter
values used are identical to those in Ref.~\cite{Luo1991}, except
$G_{\rm Ca}$ which is chosen to be $0.07$ mS cm$^{-2}$~\cite{Xie2001}.

The model equations are solved using the forward Euler method for time
evolution on a one dimensional fiber, discretized over a spatial grid of
size $L$, using a finite difference scheme for the spatial (Laplacian)
term. The values of space step $\Delta x$ and time step $\Delta t$ used
for integrating the equations are $0.0225$ cm and $0.01$ ms respectively.
The diffusion coefficient $D$ is chosen to be $0.01$ cm$^2$/msec.

%=====================================================================================================================%
\section{Results}
The key experimental results that motivate our modeling study relate
to the observation that knocking out \textit{klf15} results in a
marked reduction of $I_{\rm to}$ and that an over-expression of
\textit{klf15} leads to large increase of $I_{\rm to}$ compared to the
normal range of variation seen over the course of a day, both
increasing the susceptibility to cardiac arrhythmia. While the former
mutation showed a marked increase in the susceptibility of ventricular
arrhythmia on being subjected to programmed electrical stimulation of
the heart, the latter mutation exhibited spontaneous occurrence of
ventricular arrhythmia and resulted in a high mortality rate. In our
model, we use the closest analog of $I_{\rm to}$, viz., the  
time-dependent potassium current $I_{\rm K}$
and study the effect of increasing or
decreasing the corresponding channel conductance $G_{\rm K}$.

\begin{figure}
\begin{center}
\includegraphics[width=\textwidth]{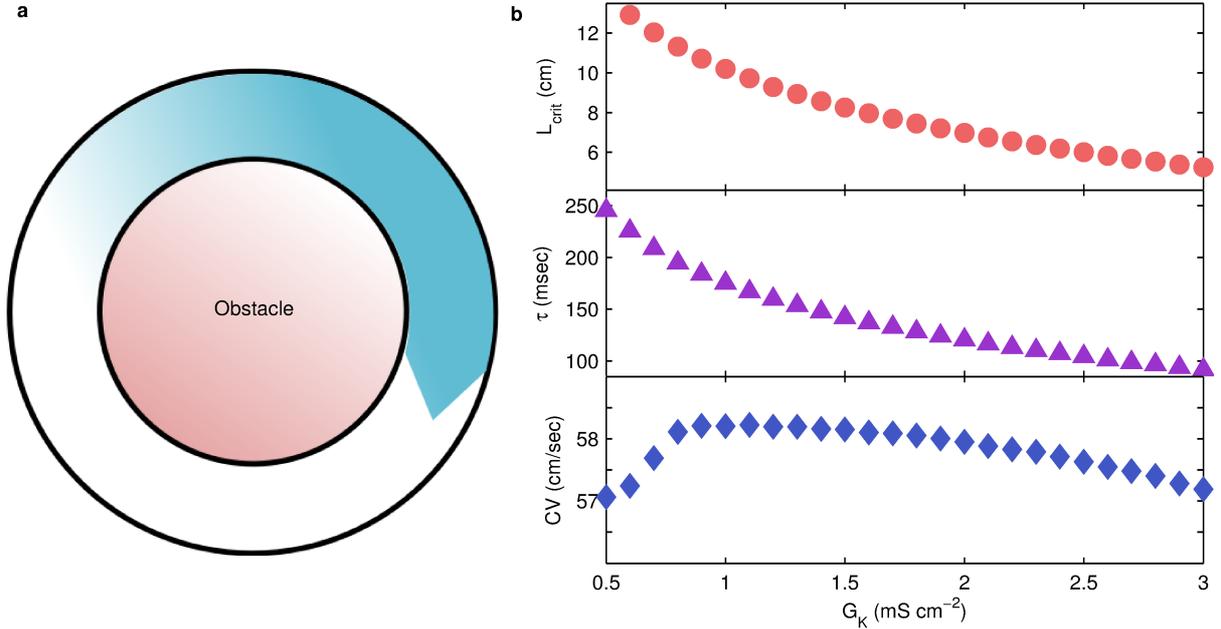}
\end{center}
\caption{Potential mechanism for genesis of pathological situations in
cardiac tissue with a high potassium ion channel conductance $G_{\rm K}$. 
(a) Schematic diagram showing the ring topology of a reentry circuit 
which describes the motion of the spiral wave around an obstacle. The 
wavefront of the excitation propagating around the ring is indicated by the
arrowhead. The color intensity represents the level of refractoriness with white
regions indicating completely recovered tissue.
(b) Variation in dynamical properties of reentrant wave propagation
as a function of $G_{\rm K}$ in a ring of critical size, i.e., the
minimum perimeter length required for sustained wave activity. 
The electrical activity of each cell is described by the LRI model.
Unidirectional wave propagation is set up using a special initial
condition. The critical ring size $L_{\rm crit}$ decreases
monotonically on increasing $G_{\rm K}$ [top], as does the period of reentry
$\tau$, i.e., the time required by the wave to complete one
circuit [center]. The conduction velocity $CV$ of the
wavefront [bottom] remains approximately constant, varying over a very
narrow range of values, as $G_{\rm K}$ is changed.
These results show that for high
$G_{\rm K}$ the critical ring size reduces, allowing reentry to occur
at higher frequency (i.e., smaller $\tau$), which could
potentially lead to wave breakup away from the reentry circuit.}
\label{fig1}
\end{figure}

Fig.~\ref{fig1} shows the role of $G_{\rm K}$ in setting up a stable
reentrant circuit. Typically, such circuits can be organized around a
zone of inexcitable or partially excitable region of cardiac tissue
(anatomical reentry) or even a transiently inactive zone which
prevents the passage of excitation wavefront across it (functional
reentry). As the wavefront of excitation goes around it, stable
reentry can occur if the circuit is long enough such that the tissue
recovers by the time the front completes the circuit. Such circuits
act as sources for persistent high-frequency stimulation that compete
with the normal rhythmic activity initiated by the sinus node. Note
that the period of the waves generated around this circuit is governed
by the circuit length which is determined by obstacle size (for anatomical
reentry) or refractory period of tissue (for functional reentry).
Focusing on the region immediately surrounding the circuit, we study
the simplified system of a one-dimensional reentrant circuit
[Fig.~\ref{fig1}~(a)]. We observe that increasing $G_{\rm K}$ results
in stable reentry being possible for circuits of smaller lengths, and
hence of much higher frequency. Stimulating the
tissue surrounding the circuit at higher frequencies will result in
waves that are more likely to result in front breakup when they travel to other regions of
the tissue~\cite{Panfilov1993,Fenton2002,Sridhar2010}. Hence
we suggest that the propensity of
life-threatening arrhythmia will be enhanced for higher $G_{\rm K}$.
Note that this is analogous to the situation corresponding to enhanced
$I_{\rm to}$ resulting from an over-expression of \textit{klf15}.

\begin{figure}
\begin{center}
\includegraphics[width=\textwidth]{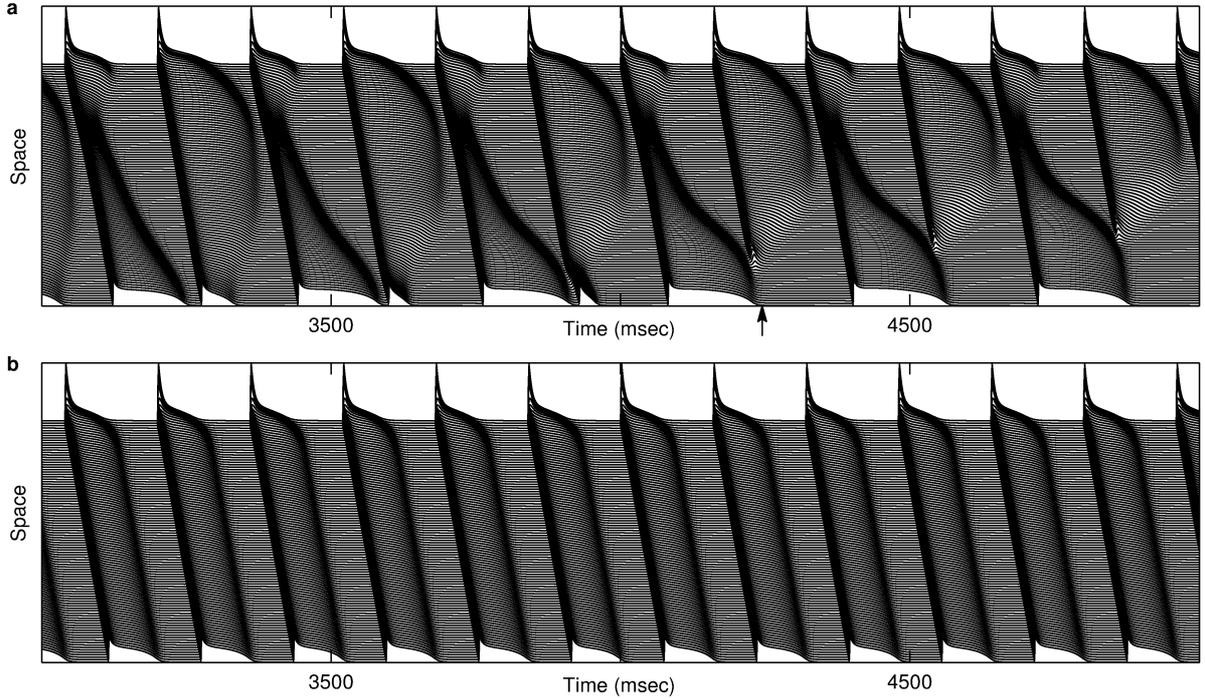}
\end{center}
\caption{Potential mechanism for genesis of pathological situations in
cardiac tissue with a low potassium ion channel conductance $G_{\rm
K}$. The dynamical response of a one-dimensional chain of cells
subjected to external periodic stimulation (pacing) with a period of
$T=160$ ms at the top end is
shown for two different values of (a) $G_{\rm K} = 0.7$ and (b) $0.9$ mS
cm$^{-2}$.
The electrical activity of each cell is described by the LRI
model. For the higher value of $G_{\rm K}$ (b), we observe a 1:1
periodic response, resulting in a train of identical action
potentials. However, for the lower value of $G_{\rm K}$ (a), pacing at
the same frequency results
in alternating action potentials varying in duration (alternans).
Moreover the pattern of alternans varies across space (discordant).
The degree of variation in the action potential duration continues to
increase over time, until a wavefront encounters a refractory region
behind the preceding wave resulting in a conduction block (see arrow).
These results show that for low $G_{\rm K}$, periodic activation of
cardiac tissue can occasionally cause
propagating fronts to be spontaneously blocked, which may then lead to
the formation of spiral waves and their breakup.
}
\label{fig2}
\end{figure}

Fig.~\ref{fig2} shows an alternative scenario where the role of
decreasing $G_{\rm K}$ in promoting arrhythmia is examined. 
This comes about through the mechanism of {\em alternans}, where periodic
stimulation of excitable media at a sufficiently high frequency
results in an alternating succession of strong and weak responses. In
cardiac tissue, this can be observed as variations in the action potential
duration (APD), such that long and short APD pulses alternate.
Such alternans can appear even at the scale of a single cell as an
outcome of the nonlinear restitution property (the functional relation
between APD and the interval between successive stimuli).
In combination with conduction velocity dispersion (i.e.,
dependence of the front propagation speed on the extent of recovery
of the medium), restitution results in an additional spatial
variation in the alternans~\cite{Sridhar2015}. Such a
variation is manifested in the form of long-short APD alternans in
certain regions co-occurring with short-long alternans in other
regions, a phenomenon termed as {\em discordant alternans}.

We study the pacing of
a one-dimensional fiber as a simple approximation of
cardiac tissue subjected to periodic electrical stimulation. 
%(as in the experimental system). 
We note that at lower values of $G_{\rm K}$ the
system is much more likely to show enhanced discordant alternans that
eventually results in conduction block far from the pacing site. Note
that in a higher-dimensional system this will result in wave-break as a
block occurs in a section of the wavefront. The broken front can then
form a reentrant circuit, resulting in arrhythmia.
%=====================================================================================================================%
\section{Discussion}
In this paper we have computationally investigated the possible mechanism underlying the recently discovered close statistical
relationship between occurrence of sudden cardiac death with the time of day. 
Essentially this involves understanding how cardiac activity is influenced by the circadian
clock, i.e., temporal oscillations in physiological activity with a
period close to $24$ hours and synchronized with the day-night cycle.
As noted in our paper, although studies have identified the genetic basis of circadian rhythm
at the intracellular level, the mechanisms by which they influence
cardiac pathologies are yet to be fully understood.
We focus on evidence suggesting that the key to the circadian-cardiac relation may be provided by the diurnal variations in the conductance
properties of ion channel proteins that govern the excitation dynamics of 
cardiac cells. 
Thus, we investigate the relationship between the circadian
rhythm as manifested in modulations of ion channel properties and the
susceptibility to cardiac arrhythmias by
using a mathematical model describing the electrical activity in
ventricular tissue. We show that changes in the channel conductance
that lead to extreme values for the duration of action potentials in cardiac
cells can result either in abnormally high-frequency reentrant
activity or spontaneous conduction block of excitation waves.
Both phenomena increase the likelihood of wavebreaks that are known to
initiate potentially life-threatening arrhythmias. Thus, disruptive
cardiac excitation dynamics are most likely to occur in time-intervals
of the day-night cycle during which the channel properties are
closest to these extreme values, providing
an intriguing relation between circadian rhythms and cardiac
pathologies.
For simplicity, here we have examined the spatially homogeneous
situation where every element in an excitable medium undergoes
similar temporal variation in the parameter controlling
repolarization. In real physiological situations, however, there may
be considerable heterogeneity in the system - e.g., in terms of the
repolarization dynamics of different regions as reflected in
potassium channel properties. Spatial variations in the recovery
characteristics are known to modulate the propagation speed of excitation
wavebacks, which can result in
wavebreaks~\cite{Breuer2005,Sridhar2010a}.
Furthermore, we have used one-dimensional models that 
consider rings or chains of cardiac cells. However, even such simple models can capture the
essential properties of excitation dynamics that we investigate here. In future studies, we plan
to show that the results obtained here under simplifying setting can be reproduced even for 
heterogeneous systems in two- or three-dimensional spatial geometries. 

\section*{Acknowledgements}
We would like to thank Gautam I. Menon for first drawing our attention to
recent experimental work relating circadian rhythms to cardiac
arrhythmia. This research was supported in part by the IMSc
Complex Systems (XII Plan) Project funded by the Department of
Atomic Energy, Government of India.

%=====================================================================================================================%


\section*{References}

\begin{thebibliography}{9}
\bibitem{Hennekens1998}
Hennekens C H 1998 
%Increasing burden of cardiovascular disease. 
\textit{Circulation} \textbf{97} 1095.

\bibitem{Heron2013}
Heron M 2013 
%Deaths: Leading causes for 2010. 
\textit{National Vital Statistics Reports} \textbf{62} 1.

\bibitem{Hatch2011} Hatch F, Lancaster M K and Jones S A 2011 \textit{Expert Rev. Cardiovasc. Ther.} \textbf{1059} 67.

\bibitem{LaraPezzi2015}
Lara-Pezzi E, Dopazo A and Manzanares M 2015 \textit{Dis. Model. Mech.} \textbf{5} 434-43.
%Understanding cardiovascular disease: a journey through the genome
%(and what we found there)

\bibitem{Mozaffarin2008} Mozaffarian D, Wilson P W and Kannel W B 2008 \textit{Circulation} \textbf{117} 3031-8.
%Beyond established and novel risk factors: lifestyle risk factors for
%cardiovascular disease

\bibitem{Dunlap1999} Dunlap J C 1999 \textit{Cell} \textbf{96} 271-90.

\bibitem{Edery2000} Edery I 2000 \textit{Physiol. Genomics} \textbf{3} 59-74.

\bibitem{Balsalobre1998} Balsalobre A, Damiola F and Schibler U 1998 \textit{Cell} \textbf{93} 929-37.

\bibitem{Durgan2005} Durgan D~J, Hotze M~A, Tomlin T~M, Egbejimi O, Graveleau C, Abel E~D, Shaw C~A, Bray M~S, Hardin P~E and Young M~E 2005 \textit{Am. J. Physiol. Heart Circ. Physiol.} \textbf{289} H1530-41.

\bibitem{Pittendrigh1993} 
Pittendrigh C~S 1993
%Temporal organization: Reflections of a Darwinian clock-watcher. 
\textit{Annu. Rev. Physiol.} \textbf{55} 17-54.

\bibitem{Ouyang1998}
Ouyang Y, Andersson C~R, Kondo T, Golden S~S and Johnson C~H 1998
%Resonating circadian clocks enhance fitness in cyanobacteria. 
\textit{Proc. Natl. Acad. Sci. USA} \textbf{95} 8660-4.

\bibitem{Paranjpe2005}
Paranjpe D~A and Sharma V~K 2005
%Evolution of temporal order in living organisms. 
\textit{J. Circadian Rhythms} \textbf{3} 7.

\bibitem{Sharma2005}
Sharma V~K and Chandrashekaran M~K 2005 
%Zeitgebers (time cues) for biological clocks. 
\textit{Curr. Sci.} \textbf{89} 1136-46.

\bibitem{Gachon2004} Gachon F, Nagoshi E, Brown S~A, Ripperger J~A and Schibler U 2004 \textit{Chromosoma} \textbf{113} 103-12.

\bibitem{Gekakis1998} Gekakis N, Staknis D, Nguyen H B, Davis F~C, Wilsbacher L~D, King D~P, Takahashi J~S and Weitz C J 1998 \textit{Science} \textbf{280} 1564-9.

\bibitem{Hogenesch1998} Hogenesch J B, Gu Y Z, Jain S and Bradfield C A 1998 \textit{Proc. Natl. Acad. Sci. USA} \textbf{95} 5474-9.

\bibitem{Kloss1998} Kloss B, Price J L, Saez L, Blau J, Rothenfluh A, Wesley C S and Young M W 1998 \textit{Cell} \textbf{94} 97-107.

\bibitem{Curtis2004} Curtis A M, Seo S B, Westgate E J, Rudic R D, Smyth E M, Chakravarti D, FitzGerald G A and McNamara P 2004 \textit{J. Biol. Chem.} \textbf{279} 7091-1097.

\bibitem{Etchegaray2003} Etchegaray J P, Lee C, Wade P A, and Reppert S M 2003 \textit{Nature} \textbf{421} 177-82.

\bibitem{Shearman2000} Shearman L P \textit{et al} 2000 \textit{Science} \textbf{288} 1013-9.

\bibitem{Cermakian2000} Cermakian N and Sassone-Corsi P 2000 \textit{Nat. Rev. Mol. Cell Biol.} \textbf{1} 59-67.

\bibitem{Hirota2004} Hirota T and Fukada Y 2004 \textit{Zoolog. Sci.} \textbf{21} 359-68.

\bibitem{Berson2003} Berson D M 2003 \textit{Trends Neurosci.} \textbf{26} 314-20.

\bibitem{Brown2002} Brown S A, Zumbrunn G, Fleury-Olela F, Preitner N, and Schibler U 2002 \textit{Curr. Biol.} \textbf{12} 1574-83.

\bibitem{Schroeder2011} Schroeder A, Loh D H, Jordan M C, Roos K P and Colwell C S 2011 \textit{Am. J. Physiol. Heart Circ. Physiol.} \textbf{300} H241-50.

\bibitem{Muller1987} Muller J E, Ludmer P L, Willich S N, Tofler G H, Aylmer G, Klangos I and Stone P H 1987 \textit{Circulation} \textbf{75} 131-8.

\bibitem{Jeyaraj2012} Jeyaraj D \textit{et al} 2012 \textit{Nature} \textbf{483} 96-9.

\bibitem{Schroder2013} Schroder E A, Lefta M, Zhang X, Bartos D C, Feng H Z, Zhao Y, Patwardhan A, Jin J P, Esser K A and Delisle B P 2013 \textit{Am. J. Physiol. Cell Physiol.} \textbf{304} C954-65.

\bibitem{Jalife1999} Jalife J, Anumonwo J M, Delmar M and Davidenko J
M 1999 \textit{Basic Cardiac Electrophysiology for the Clinician}
(Armonk: Futura Publishing).

\bibitem{Luo1991} Luo C and Rudy Y 1991 \textit{Circ. Res.} \textbf{68} 1501-26.

\bibitem{Hodgkin1952} Hodgkin A L and Huxley A F 1952 \textit{J. Physiol.} \textbf{117} 500-44.

\bibitem{Xie2001}
Xie F, Qu Z, Garfinkel A and Weiss J~N 2001 \textit{Am. J.
Physiol. Heart Circ. Physiol.} \textbf{280} H535-45.

\bibitem{Panfilov1993}
Panfilov A V and Keener J P 1993 \textit{J. Theor. Biol.}
\textbf{163}, 439-48.

\bibitem{Fenton2002}
Fenton F H, Cherry E M, Hastings H M and Evans S J 2002
\textit{Chaos} \textbf{12}, 852-92.

\bibitem{Sridhar2010}
Sridhar S, Sinha S and Panfilov A V 2010 \textit{Phys. Rev. E}
\textbf{82}, 051908.

\bibitem{Sridhar2015}
Sinha S and Sridhar S 2015 \textit{Patterns in Excitable Media: Genesis,
Dynamics, and Control} (Boca Raton, FL: CRC Press). 

\bibitem{Breuer2005}
Breuer J and Sinha S 2005 \textit{Pramana} \textbf{64}, 553-62.

\bibitem{Sridhar2010a}
Sridhar S and Sinha S 2010 \textit{EPL} \textbf{92}, 60006.

\end{thebibliography}
\end{document}